\documentclass[twocolumn,letter]{jpsj2} 

\def\runauthor{T. \textsc{Sasaki} {\it et al.}}

\title{%
Superconducting Properties under Magnetic Field in Na$_{0.35}$CoO$_{2}{\cdot}1.3$H$_{2}$O Single Crystal
}

\author{%
Takahiko~\textsc{Sasaki}\thanks{E-mail address: takahiko@imr.tohoku.ac.jp}, Petre~\textsc{Badica}, Naoki~\textsc{Yoneyama}, Kazuyoshi~\textsc{Yamada}, Kazumasa~\textsc{Togano} and Norio~\textsc{Kobayashi}
}

\inst{%
Institute for Materials Research, Tohoku University, Katahira 2-1-1, Aoba-ku, Sendai 980-8577, Japan
}

\recdate{}

\abst{%
We report the in-plane resistivity and magnetic susceptibility of the layered cobalt oxide Na$_{0.35}$CoO$_{2}{\cdot}1.3$H$_{2}$O single crystal.  The temperature dependence of the resistivity shows metallic behavior from room temperature to the superconducting transition temperature $T_{c}$ of 4.5 K.  Sharp resistive transition, zero resistivity and almost perfect superconducting volume fraction below $T_{c}$ indicate the good quality and the bulk superconductivity of the single crystal.  The upper critical field $H_{c2}$ and the coherence length $\xi$ are obtained from the resistive transitions in magnetic field parallel to the $c$-axis and the $ab$-plane.  The anisotropy of $\xi$, $\xi_{ab}$/$\xi_{c} =$ 12 nm/1.3 nm $\simeq$ 9.2, suggests that this material is considered to be an anisotropic three dimensional superconductor.  In the field parallel to the $ab$-plane, $H_{c2}$ seems to be suppressed to the value of Pauli paramagnetic limit. It may indicate the spin singlet superconductivity in the cobalt oxide.
}

\kword{%
Na$_{0.35}$CoO$_{2}{\cdot}1.3$H$_{2}$O, superconductivity, upper critical field, anisotropy
}

\begin{document}
\maketitle

The superconductivity in the layered cobalt oxide Na$_{0.35}$CoO$_{2}{\cdot}1.3$H$_{2}$O found by Takada {\it et al.}\cite{Takada} has attracted much attention.  
In particular, unconventional superconductivity has been expected in this compound because of the unique crystal structure in the CoO$_{2}$ layer although the superconducting transition temperature $T_{c} \simeq$ 5 K is one order of magnitude smaller than that in High-$T_{c}$ copper oxide superconductors.  
The CoO$_{2}$ layer is formed by the hexagonal arrangement of the Co ions, which is in contrast to the tetragonal arrangement of the Cu ions in the copper oxides.  
A deficiency of the sodium atom leads to dope 0.65 holes into the layer of the insulating low-spin ($S = 0$) state of Co$^{+3}$.  
This hole doping results in 0.35 electrons doped in the triangular lattice consisting of $S = 1/2$ of Co$^{+4}$.  
Such triangular magnetic lattice is regarded as having magnetic frustration and one may expect that unconventional superconductivity is realized on the basis of such frustration.  
Then just after finding the superconductivity in the present material, many theoretical proposals of possible unconventional superconductivity have been presented.\cite{Baskaran,Kumar,Ogata,Tanaka}  
The experimental investigation on the superconductivity, however, has not progressed so much because of difficulty of getting the single crystal.  
It is highly desirable to obtain the basic superconducting parameters in magnetic field, for example, the upper critical field $H_{c2}$, the coherence length $\xi$, and those anisotropy.  
These basic material parameters must be important to understand NMR and NQR results reported so far,\cite{Kobayashi,Ishida} which should give crucial information about the pairing symmetry and the gap anisotropy.  

In this letter, the basic superconducting parameters such as $H_{c2}$, $\xi$, and those anisotropy of the superconducting cobalt oxide are reported on the basis of the resistivity measurements in the single crystal.  

Single crystals of Na$_{0.7}$CoO$_{2}$ used in this study was grown by floating zone method.  
The deintercalation of Na and hydration process to obtain the superconductivity
is similar way reported by Takada {\it et al.}\cite{Takada}  
The detail of the sample preparation and the characterization will be published elsewhere.\cite{Badica}
The superconducting crystals were obtained as Na$_{0.35}$CoO$_{2}{\cdot}1.3$H$_{2}$O.  
The typical size of the single crystal is 1 $\times$ 0.6 $\times$ 0.02 mm$^{3}$.  
The in-plane resistivity was measured by standard dc four probe method.  
In order to make the electrical contact to the sample, the gold wires (10 $\mu$m$\phi$) were attached with silver paste which dried for 30 minutes at a few C$^{\circ}$.  
The samples were fixed to the single-axis rotatable holder for resistivity measurements in magnetic fields and cooled quickly below ice point because of avoiding the partial dehydration from the sample.  
We paid attention to handle the samples in cold condition.  
For instance, the samples were placed on the aluminum plate cooled by ice during the preparation of the electrical contact under microscope.  
The magnetic susceptibility was measured with a SQUID magnetometer (Quantum Design MPMS-XL5).  


\begin{figure}[t]
\begin{center}
\includegraphics[viewport=2cm 7cm 19cm 23cm,clip,width=0.8\linewidth]{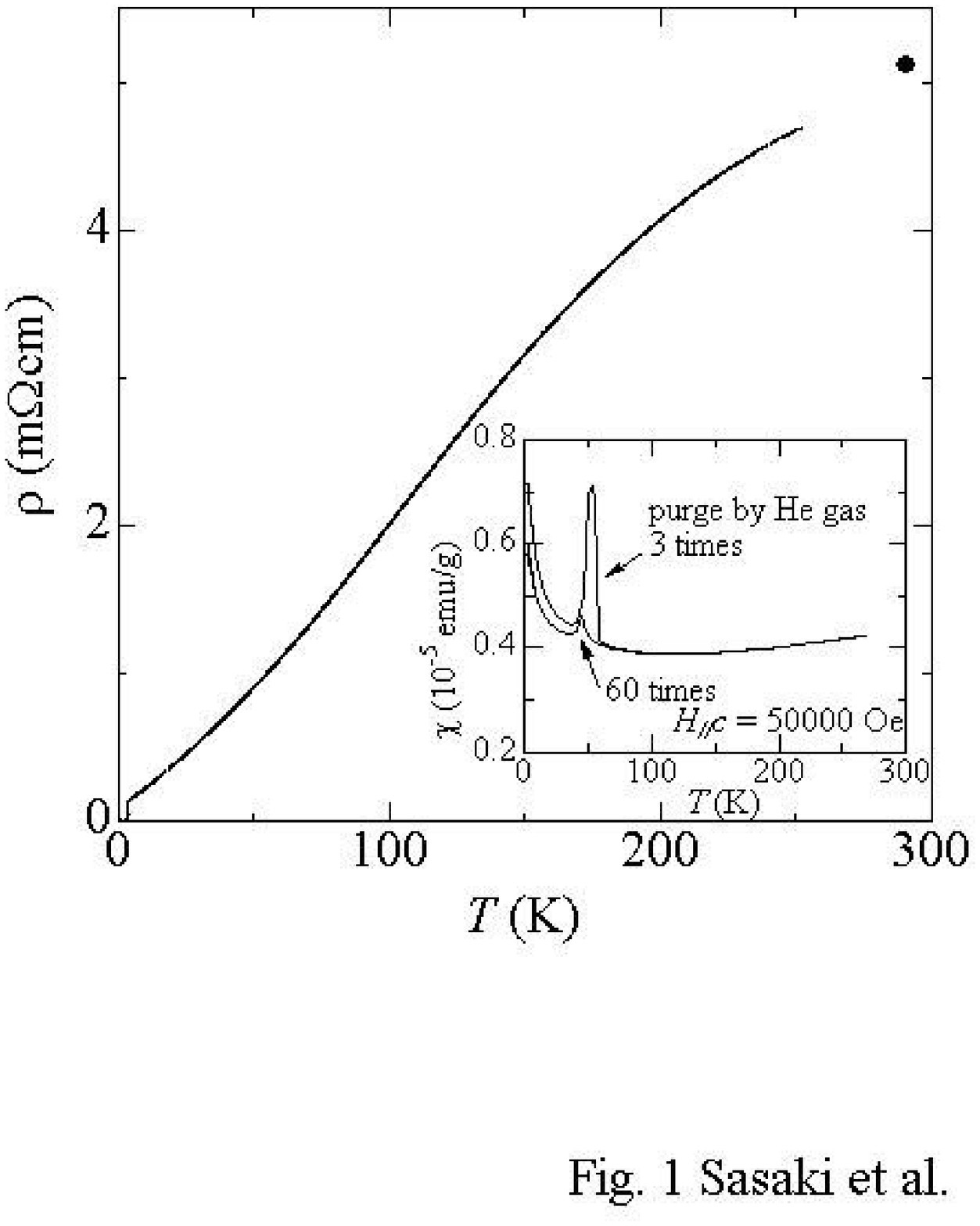}
\end{center}
\caption{Temperature dependence of the in-plane resistivity of Na$_{0.35}$CoO$_{2}{\cdot}1.3$H$_{2}$O single crystal.  Filled circle indicates the data point at 290 K. In order to cool the sample as rapidly as possible, no data between 250 and 290 K is available. Inset shows the magnetic susceptibility in $H =$ 50000 Oe parallel to the $c$-axis. The sample space is purged by helium gas at room temperature three (upper curve) and sixty times (lower curve).}
\label{fig1}
\end{figure}

Figure 1 shows the temperature dependence of the in-plane resistivity of the single crystal sample in the cooling process.  
The absolute value of the in-plane resistivity is somewhat larger than the reported value\cite{Jin,Chou} of 1 -- 2 m$\Omega$cm at room temperature.  
The present resistivity may have an ambiguity because of the irregular shape and easily cleaved nature of the layered crystal.\cite{Badica}  
The data between 250 K and 290 K is lacked due to the quick cooling from room temperature.  
The resistivity shows monotonic decrease from 290 K down to $T_{c}$.  
The quality of the single crystal seems to be good considering large residual resistivity ration $\rho$(290 K)/$\rho$(5 K) $\simeq$ 38.  
The superconducting transition appears following such metallic temperature dependence of the resistivity.  
Such metallic behavior in the single crystal sample is different from the previous report by Jin {\it et al.},\cite{Jin} but is similar to another report by Chou {\it et al.}\cite{Chou} although both groups use single crystals. 
The former results have shown a large resistance anomaly around 50 K. 
The origin of the difference is not known at present, but the metallic behavior in the in-plane resistivity observed by Chou {\it et al.} and us seems to be intrinsic nature because no anomaly in magnetic susceptibility mentioned latter and specific heat has been reported at the temperature.\cite{Jin}  

The temperature dependence of the magnetic susceptibility $\chi$ in another single crystal from the same batch is shown in the inset.  
Around 45 K, an anomaly appears as a cusp of $\chi$, which has been reported in the single crystal sample by Chou {\it et al.},\cite{Chou} but has not been observed in the powder sample.\cite{Ishida,Jin,Sakurai}
The origin of the anomaly is likely to be the residual oxygen on the surface or in the intercalated water because the large number of purging the sample space of the magnetometer by helium gas at room temperature strongly suppresses the anomaly.  
The Curie-Weiss like enhancement of $\chi$ at low temperature is reduced also by such sufficient gas exchange.  
The difference of the Curie-Weiss like enhancement by purging appears only below the temperature of the cusp anomaly. 
Therefore the anomaly of $\chi$ around 45 K and some part of the Curie-Weiss like enhancement of $\chi$ can be considered to come from the extrinsic origin which is probably due to the residual oxygen.
It is noted, however, that the Curie-Weiss like enhancement becomes remaining even after sufficient number of purging. 
It suggests that some part of the enhancement is free from the residual oxygen and may result from a magnetic fluctuation origin.\cite{Ishida}

\begin{figure}[t]
\begin{center}
\includegraphics[viewport=0cm 4cm 18cm 25cm,clip,width=0.8\linewidth]{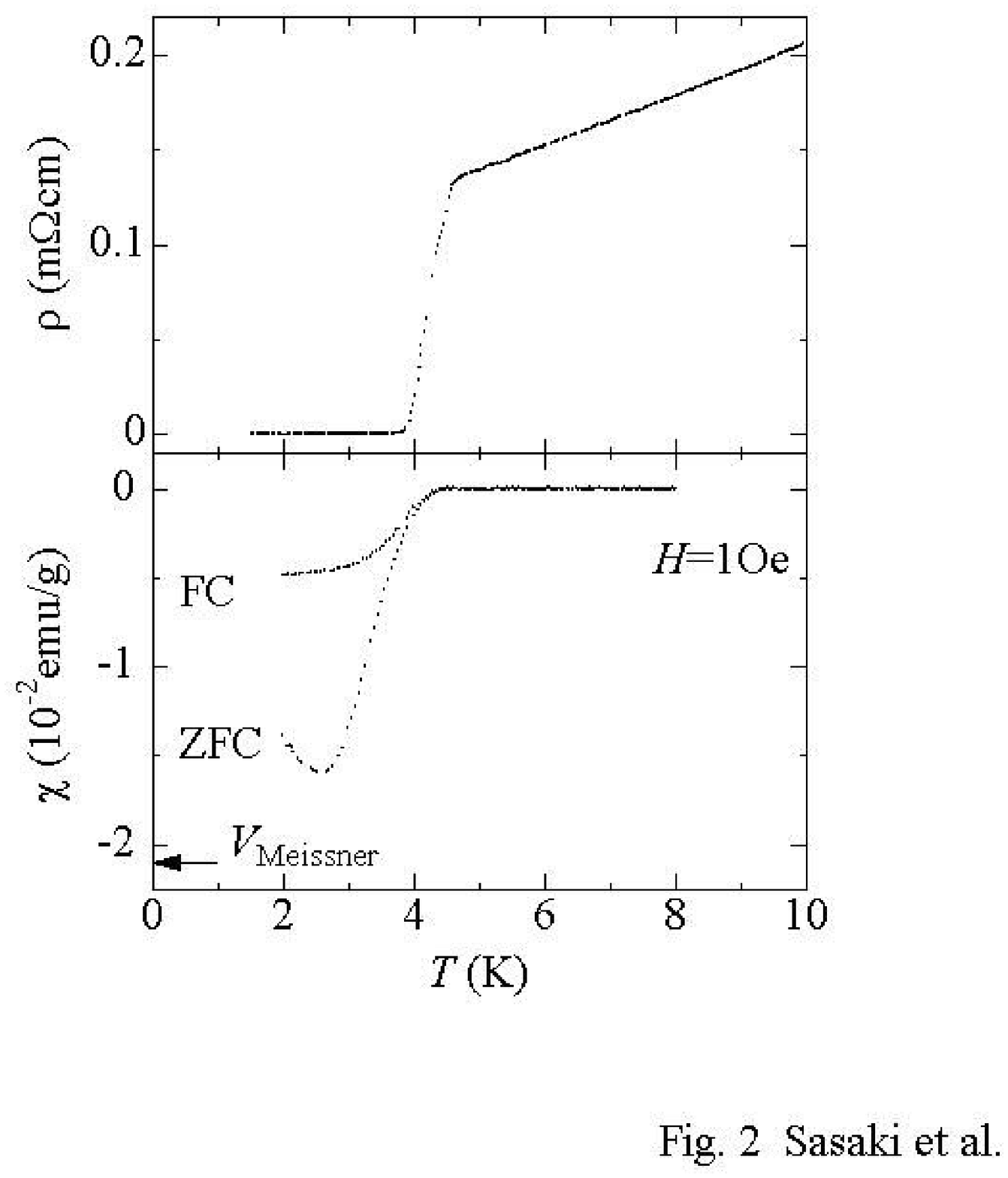}
\end{center}
\caption{Temperature dependence of (a) the in-plane resistivity and (b) the magnetic susceptibility $\chi$ in $H =$ 1 Oe in the same single crystal of Na$_{0.35}$CoO$_{2}{\cdot}$1.3H$_{2}$O. V$_{\rm Meissner}$ denotes the full volume of the superconductivity.}
\label{fig2}
\end{figure}

Figure 2 demonstrates the superconducting transition observed by the in-plane resistivity (upper figure) in zero magnetic field and the magnetic susceptibility in $H =$1 Oe parallel to the $c$-axis.  
Both the resistivity and the magnetic susceptibility were measured on the same sample.  
The sample after measuring resistivity was transferred to the magnetometer in keeping the temperature cold below ice point as rapidly as possible.
The resistive transition starts at about 4.5 K.  
It shows zero resistivity at 3.9 K, which is smaller than the present experimental accuracy, 10$^{-3}$ of the normal resistivity value at 5 K. 
There are small structures seen in the transition curve.  
It may be due to distribution of $T_{c}$ inside the crystal which may be caused by the partial dehydration at room temperature.
The magnetic susceptibility which is corrected by the demagnetization factor using a ellipsoidal approximation decreases below about 4.5 K.  
The magnitude of the shielding signal in zero magnetic field cooling (ZFC) indicates almost 70 -- 80 \% of the superconducting fraction to the full volume V$_{\rm Meissner}$ of the crystal although the value has uncertainty coming from the ambiguity of the demagnetization factor of the planer shape of the sample.  
The sharp transition and almost full superconducting volume fraction demonstrate that the sample used here shows the bulk superconductivity and has good enough quality for evaluating the intrinsic superconducting parameters.

\begin{figure}[t]
\begin{center}
\includegraphics[viewport=0cm 5cm 19cm 23cm,clip,width=0.8\linewidth]{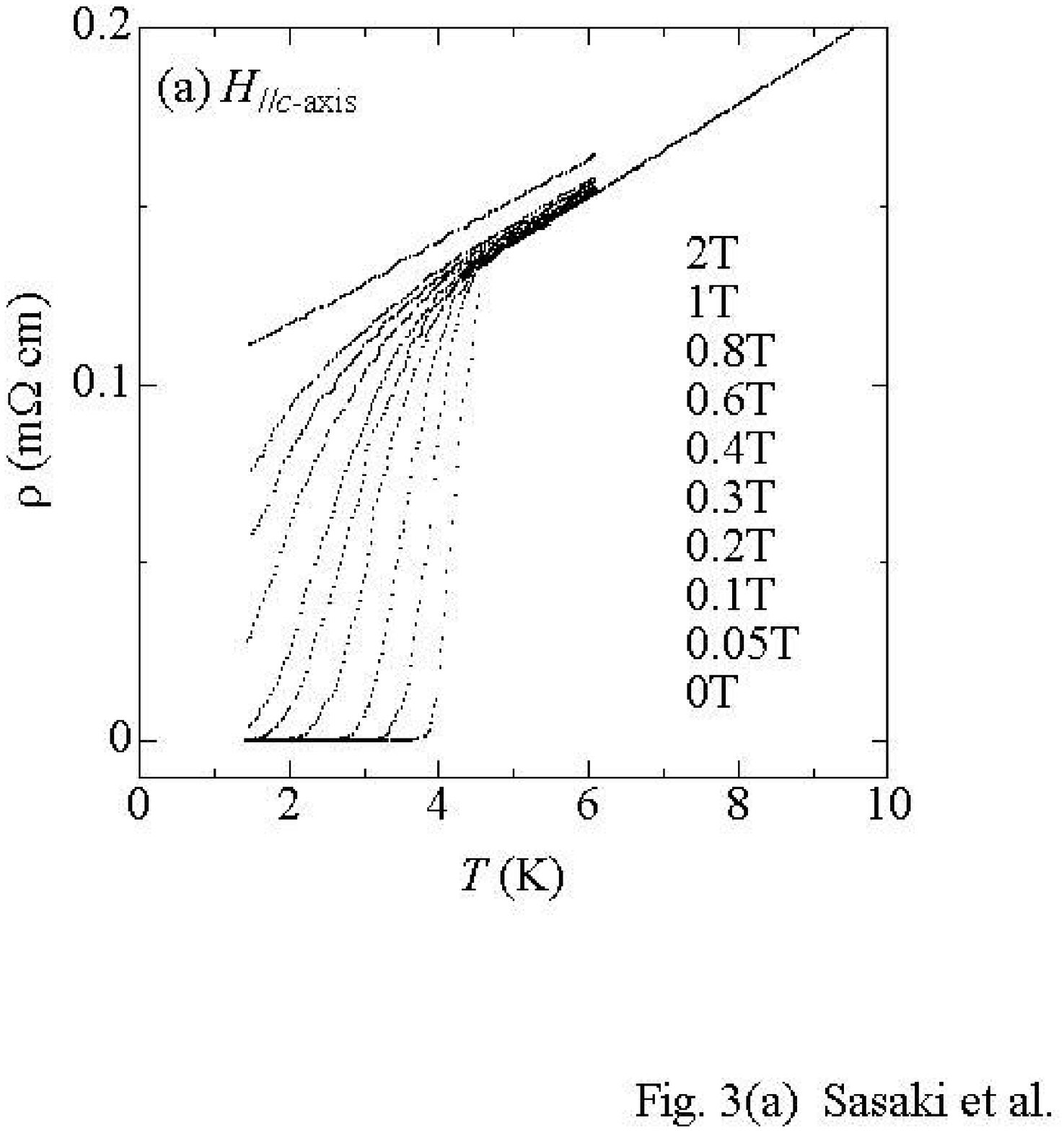}
\includegraphics[viewport=1cm 6cm 20cm 24cm,clip,width=0.8\linewidth]{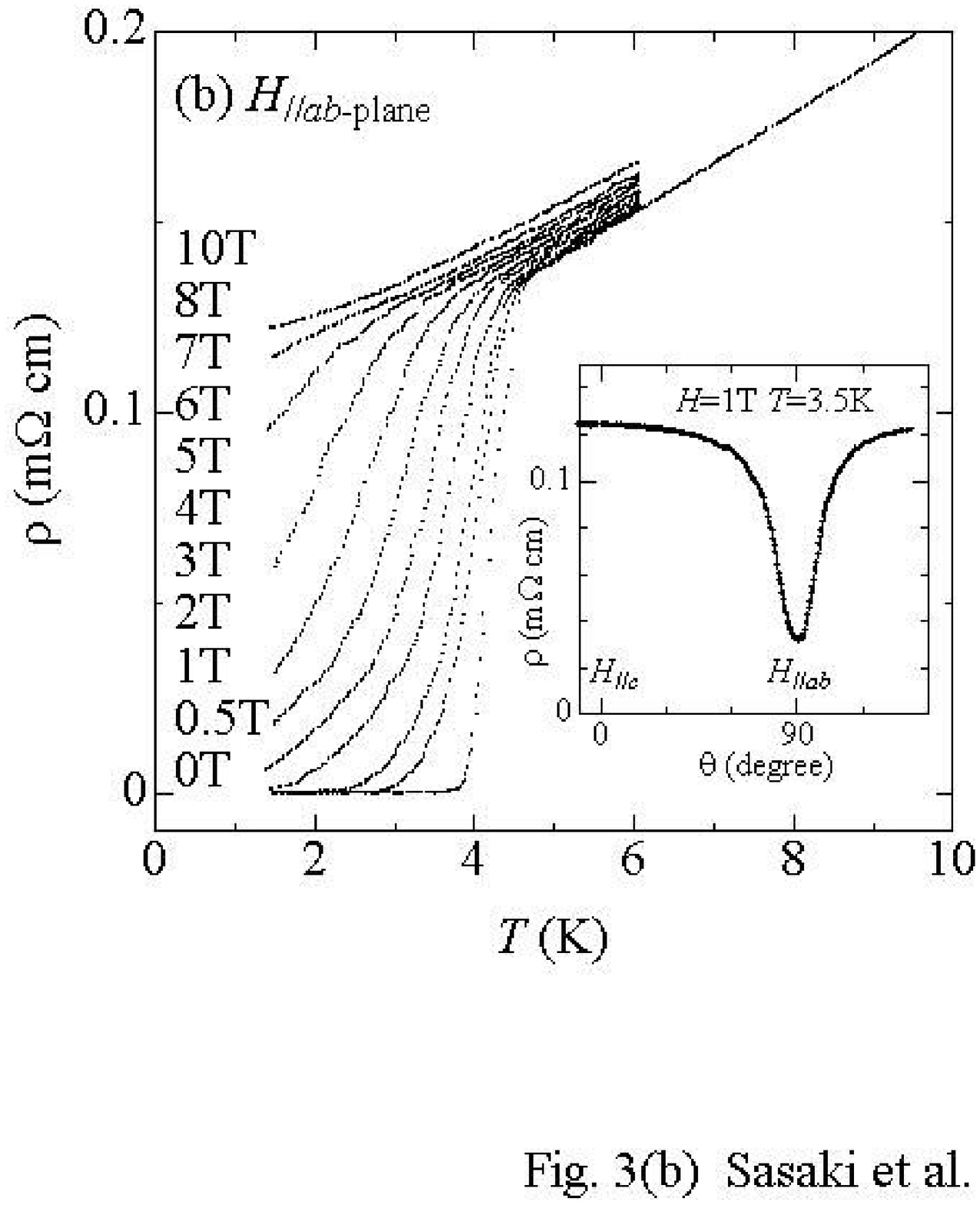}
\end{center}
\caption{Temperature dependence of the in-plane resistivity under magnetic fields parallel to (a) the $c$-axis and (b) the $ab$-plane of Na$_{0.35}$CoO$_{2}{\cdot}1.3$H$_{2}$O single crystal.  The inset shows the magnetic field direction dependence of the resistivity at $T =$ 3.5 K and in $H =$ 1 T. The magnetic field direction is denoted by the angle $\theta$ tilted from the $c$-axis to the $ab$-plane.}
\label{fig3}
\end{figure}

Figures 3(a) and 3(b) show the temperature dependence of the in-plane resistivity under magnetic fields parallel to the $c$-axis and $ab$-plane, respectively.  
In order to align the magnetic field exactly parallel to the $ab$-plane, the magnetic field direction dependence of the resistivity is measured.  
The resistivity takes its minimum in the field parallel to the $ab$-plane which is shown in the inset of Fig. 3(b).  
The in-plane current is always perpendicular to the magnetic field.  
The resistivity minimum should show sharp cusp behavior\cite{Akima} in the case of the large anisotropy expected from the large CoO$_{2}$ layer spacing expanded by the hydration.  
The shape of the minimum, however, looks to be rounded.
It is not clear at present that the reason of the round shape of the minimum is whether the anisotropy is not so large intrinsically or some experimental artifacts such as the large misalignment in setting the sample to single-axis rotator and the possible slight bending in the crystal used.  
But we notice the gradual decrease of the resistivity can be seen already in the region of $\pm$ 30 degree from the parallel direction. 
It may indicate that the intrinsic anisotropy is not so large because the artifact origin is not effective in such angle region.  

The resistive transition becomes broad in both magnetic field directions. 
Then the certain value of $H_{c2}$ is difficult to be determined from the resistive transition.  
Here we define some characteristic points in the resistive transition curves by using several criteria of the resistivity value.  
The criteria are 0.9$\rho_{n}$, 0.5$\rho_{n}$, 0.1$\rho_{n}$, and 0.001$\rho_{n}$, where $\rho_{n}$ is the normal resistivity extrapolated from higher temperature region. 
In the temperature region above 1.5 K, the temperature dependence of the normal resistivity is expected to show a $T$-linear dependence from the data in 2 T ($H \parallel$ $c$-axis) and 8 T ($H \parallel$ $ab$-plane).

\begin{figure}[t]
\begin{center}
\includegraphics[viewport=2cm 6.5cm 20cm 24cm,clip,width=0.8\linewidth]{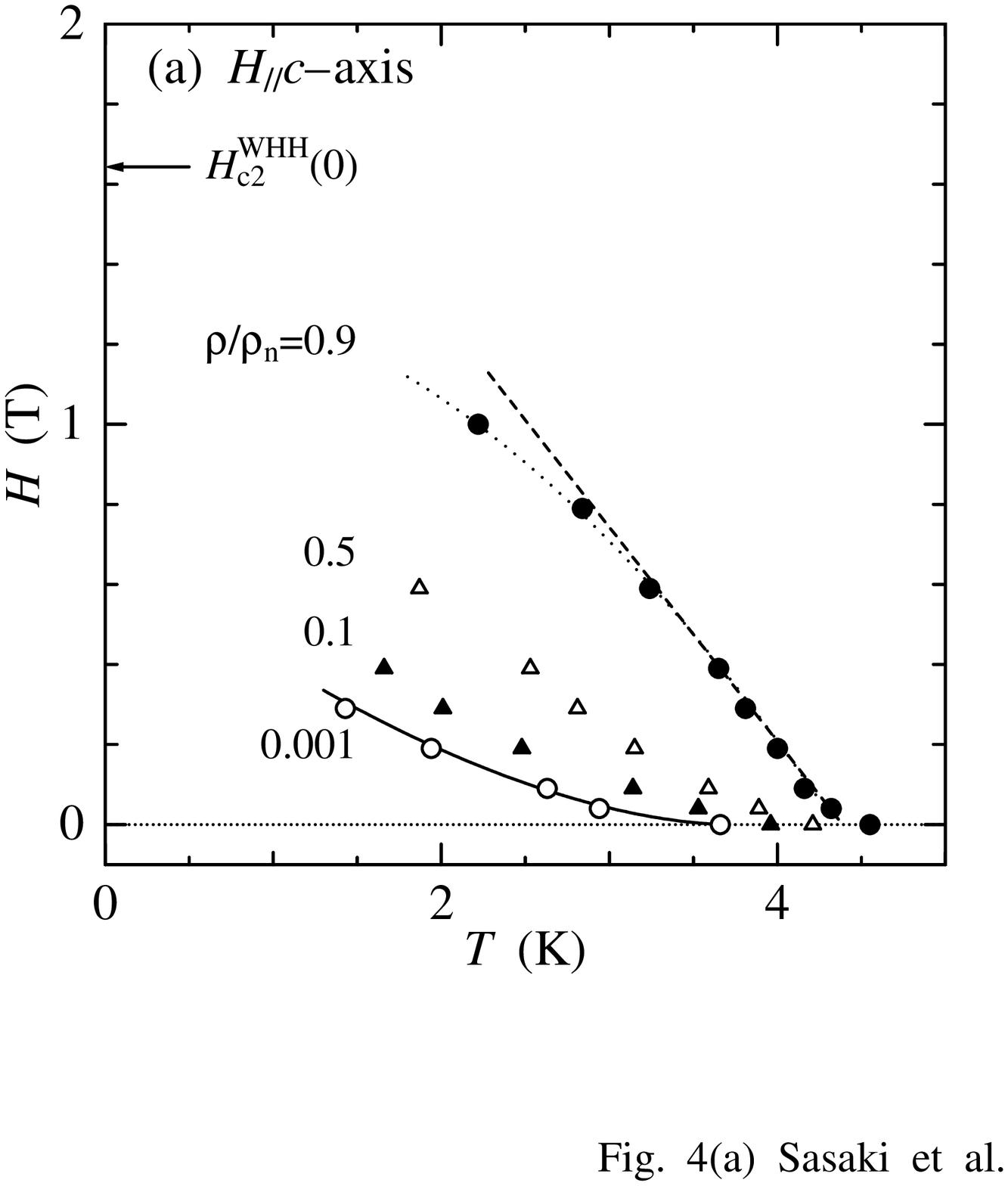}
\includegraphics[viewport=2cm 6.5cm 20cm 24cm,clip,width=0.8\linewidth]{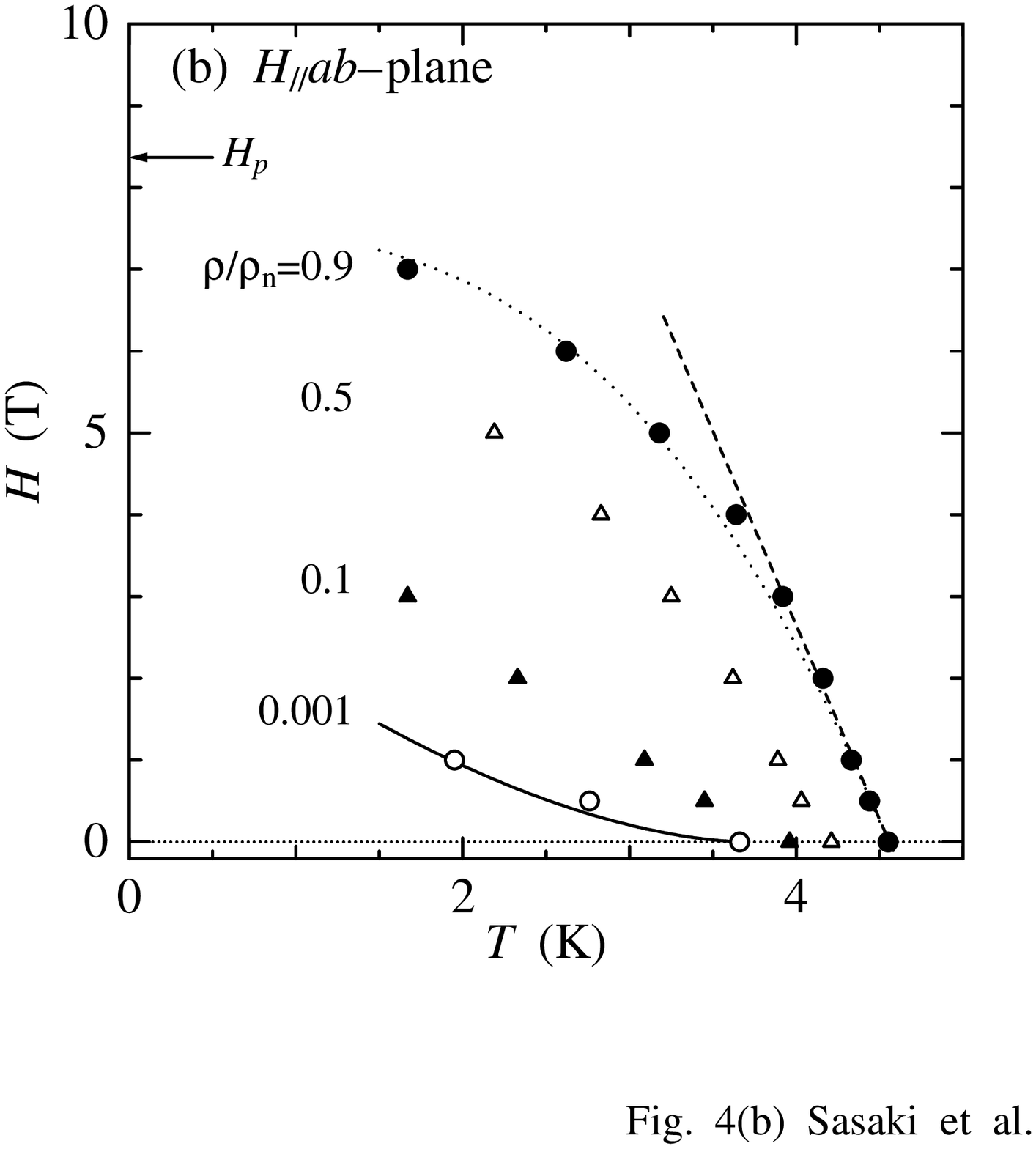}
\end{center}
\caption{Magnetic field - temperature phase diagram of Na$_{0.35}$CoO$_{2}{\cdot}1.3$H$_{2}$O in the field parallel to (a) the $c$-axis and (b) the $ab$-plane. The filled circle, open triangle, filled triangle and open circle denote the points defined by the resistivity level of 0.9$\rho_{n}$, 0.5$\rho_{n}$, 0.1$\rho_{n}$ and 0.001$\rho_{n}$, respectively.  The dotted and solid curves are guides for eyes.  The dashed lines show the initial slope of $H_{c2}$ defined by $\rho/\rho_{n} =$ 0.9 criterion. $H_{c2}^{\rm WHH}$(0) and $H_{p}$ are the upper critical field estimated by WHH and the Pauli paramagnetic limit, respectively.}
\label{fig4}
\end{figure}

Figures 4(a) and 4(b) show the magnetic field - temperature phase diagram on the resistive characteristic transition points in the magnetic field parallel to the $c$-axis and $ab$-plane, respectively.  
The filled circles, open triangles, filled triangles, and open circles denote the points defined by the several resistivity criteria of 0.9$\rho_{n}$, 0.5$\rho_{n}$, 0.1$\rho_{n}$, and 0.001$\rho_{n}$, respectively.  
The dotted and solid curves corresponding to 0.9$\rho_{n}$ and 0.001$\rho_{n}$ are guides for eyes.  
Here we use the resistive criteria 0.9$\rho_{n}$ and 0.001$\rho_{n}$ for the upper critical field $H_{c2}$ and the irreversibility field $H_{\rm irr}$ in conventional way although such criteria are not sound exactly correct for thermodynamic transitions.  
The initial slope of $H_{c2}(T)$ at $T_{c}$ shown as the dashed lines are $-0.53$  and $-4.7$ T/K in the field parallel to the $c$-axis and $ab$-plane.
Theses values of the slope are in agreement with the previous reports.\cite{Badica,Chou}
From these initial slopes, the Ginzburg - Landau coherence length $\xi$ can be calculated by using the relation $dH_{c2}^{i}/dT|_{Tc} = -(1/T_{c})(\Phi_{0}/2\pi\xi_{j}\xi_{k})$, where $\Phi_{0} = hc/2e = 2.07 \times 10^{-11}$ T$\cdot$cm$^{2}$ and $i$, $j$, and $k$ are cyclic permutations of the directions.  Here we assume the isotropic $\xi$ in the $ab$-plane, that is $\xi_{i} = \xi_{j} = \xi_{ab}$ and $\xi_{k} = \xi_{c}$.  
Following the equations the coherence lengths are calculated to be $\xi_{ab} \simeq$ 12 nm and $\xi_{c} \simeq$ 1.3 nm.  
It is noted that the interlayer coherence length $\xi_{c}$ is longer than the interlayer spacing, about 0.98 nm, of the CoO$_{2}$ layers.\cite{Takada}
Accordingly the anisotropy calculated from $\xi$ is about 9.2, which may be a lower limit value of the anisotropy considering the experimental uncertainty described before.
This moderate size of the anisotropy is almost comparable to that of the pyridine intercalated transition metal dicalcogenides ($\sim$ 20),\cite{Morris} organics ($\sim$ 10)\cite{Lang} and Sr$_{2}$RuO$_{4}$ ($\sim$ 20)\cite{Akima} and is larger than that of the alkali intercalated nitrides ($\sim$ 4).\cite{Tou}  
In comparison with the High-$T_{c}$ copper oxides, the anisotropy of the present cobalt oxide is located between the value of YB$_{2}$Cu$_{3}$O$_{7-\delta}$ and the La-based cuprates.\cite{oxide} 
One may expect larger anisotropy than the observed value because of expanding largely the interlayer spacing between the CoO$_{2}$ layers by intercalation of H$_{2}$O molecules.  
But it is much smaller than that of the Bi-based cuprates.  
The reason may be due to longer interlayer coherence length than the interlayer spacing of the CoO$_{2}$ layers.  

The upper critical field at $T =$ 0 K can be estimated from Werthamer-Helfand-Hohenberg (WHH) formula\cite{WHH} by using the initial slope, $H_{c2}^{\rm WHH}(0) = -0.693(dH_{c2}/dT|_{T_{c}})T_{c}$.  
The calculated values of $H_{c2}^{\rm WHH}(0)$ are about 1.7 T and 15 T in the field parallel to the $c$-axis and $ab$-plane, respectively.  
In the magnetic field parallel to the $c$-axis, the $H_{c2}^{c}(T)$ line looks likely to be pointing to $H_{c2}^{\rm WHH}(0)$.  
On one hand, $H_{c2}^{ab}(T)$ is strongly suppressed from $H_{c2}^{\rm WHH}(0)$ in the low temperature region.  
This suppression may arise from the singlet pair breaking by Zeeman effect, that is so-called Clogston-Chandrasekhar limit\cite{Clogston,Chandrasekhar} or Pauli paramagnetic limit.  
The value of Pauli paramagnetic limit is simply described as $H_{P} = 1.84T_{c}$.  
According to $T_{c} = 4.5$ K, $H_{P} \simeq$ 8.3 T.  
As can be seen in Fig. 4(b), the actual $H_{c2}^{ab}(T)$ line seems to be suppressed from $H_{c2}^{\rm WHH}(0)$ to $H_{\rm P}$.  
Such suppression by Zeeman effect may suggest the spin singlet superconductivity in the present cobalt oxide superconductor.  

There is large discrepancy on $H_{c2}$ value between the recent report\cite{Sakurai} in the powder sample and the present results.  
The magnetization measurements in the powder sample have been used to estimate $H_{c2}$ and the lower critical field $H_{c1}$ by Sakurai {\it et al.}\cite{Sakurai}
They reported huge $H_{c2}^{\rm WHH}(0) =$ 61 T evaluated from the initial slope $dH_{c2}/dT|_{T_{c}} =$ --19.3 T/K at $T_{c}$.  
These value of $H_{c2}^{\rm WHH}(0)$ and the initial slope are about four times larger than our results in magnetic fields parallel to the $ab$-plane.  
The reason of such large discrepancy is not known at present.
But judging from the resistivity curve in $H =$ 10 T shown in Fig. 3(b), it is naturally considered that there is no indication of the superconductivity down to 1.5 K even in the case of the magnetic field parallel to the $ab$-plane.  

In conclusion, we show the upper critical field and its anisotropy evaluated from the resistivity measurements on the single crystal sample of the layered cobalt oxide Na$_{0.35}$CoO$_{2}{\cdot}1.3$H$_{2}$O.  
The magnitude of the anisotropy and the coherence length in the interlayer direction indicate that this superconductor is classified to the anisotropic three dimensional one.  
The suppression of the upper critical field to the Pauli limiting value may suggest the spin singlet pairing of the superconductivity of this compound.  
In order to confirm the present results, lower temperature experiments on the phase diagram must be important because several unconventional features, such as multi superconducting phases, are expected in the unconventional superconductor having a triplet pairing.

The authors thank W. Koshibae, K. Kudo, T. Nishizaki and T. Tohyama for stimulating discussions.

\end{document}